\documentclass[twocolumn,preprint,3p,twocolumn]{elsarticle}
\usepackage[latin9]{inputenc}
\usepackage{amsmath}
\usepackage{amssymb}
\usepackage{graphicx}
\usepackage{esint}
\usepackage{natbib}
\biboptions{numbers,sort&compress} 

\makeatletter

\usepackage{color}\usepackage{epsfig}\usepackage{dcolumn}\usepackage{bm}\usepackage{color}

\makeatother

\begin{document}
\begin{frontmatter}

\title{Fano fingerprints of Majoranas in Kitaev dimers of superconducting adatoms}

\author{F. A. Dessotti$^{1}$, L. S. Ricco$^{1}$, Y. Marques$^{1}$, R.
S. Machado$^{1}$, L. H. Guessi$^{2}$,\\
 M. S. Figueira$^{3}$, M. de Souza$^{2}$ and A. C. Seridonio $^{1,2}$\corref{A}}

\address{$^{1}$Departamento de Física e Química, Universidade Estadual Paulista,
15385-000, Ilha Solteira, São Paulo, Brazil\\
 $^{2}$Instituto de Geociências e Ciências Exatas - IGCE, Universidade
Estadual Paulista, Departamento de Física, 13506-970, Rio Claro, São
Paulo, Brazil\\
 $^{3}$Instituto de F\'{i}sica, Universidade Federal Fluminense,
24210-340, Niterói, RJ, Brazil}

\cortext[A]{Corresponding Author: seridonio@dfq.feis.unesp.br}

\begin{abstract}
We investigate theoretically a Fano interferometer composed by STM
and AFM tips close to a Kitaev dimer of superconducting adatoms, in
which the adatom placed under the AFM tip, encloses a pair of Majorana
fermions (MFs). For the binding energy $\Delta$ of the Cooper pair
delocalized into the adatoms under the tips coincident with the tunneling
amplitude $t$ between them, namely $\Delta$ = $t$, we find that
only one MF beneath the AFM tip hybridizes with the adatom coupled
to the STM tips. As a result, a gate invariance feature emerges: the
Fano profile of the transmittance rises as an invariant quantity depending
upon the STM tips Fermi energy, due to the symmetric swap in the gate
potential of the AFM tip.
\end{abstract}

\end{frontmatter}

\section{Introduction}

\label{sec1}

The physicist Ettore Majorana proposed in the field of high-energy
Physics the existence of peculiar fermions that constitute their own
antiparticles. In the context of condensed matter Physics, these fermions
are Majorana quasiparticles \citep{key-24,Franz}. From the quantum
computing perspective, two Majorana fermions (MFs) can compose a regular
fermion acting as a protected qubit, which is indeed decoupled from
the host environment and free of the decoherence effect. As a result,
the quest for setups supporting MFs has attracted broad interest from
the communities of researchers in the fields of quantum information
and transport \citep{key-36,key-37,key-377}, since the qubit made
by the coupled MFs appear only in the topological phase. Noteworthy,
the Kitaev chain within such a phase \citep{key-333} is considered
the most promising candidate to this end as the aftermath of the emerging
$p$-wave and spinless superconductivity. Indeed, in Kitaev's setup,
MFs appear as zero-energy modes attached to the edges of the chain.

\begin{figure}[!htb]
\begin{centering}
\includegraphics[width=0.5\textwidth,height=0.5\textheight]{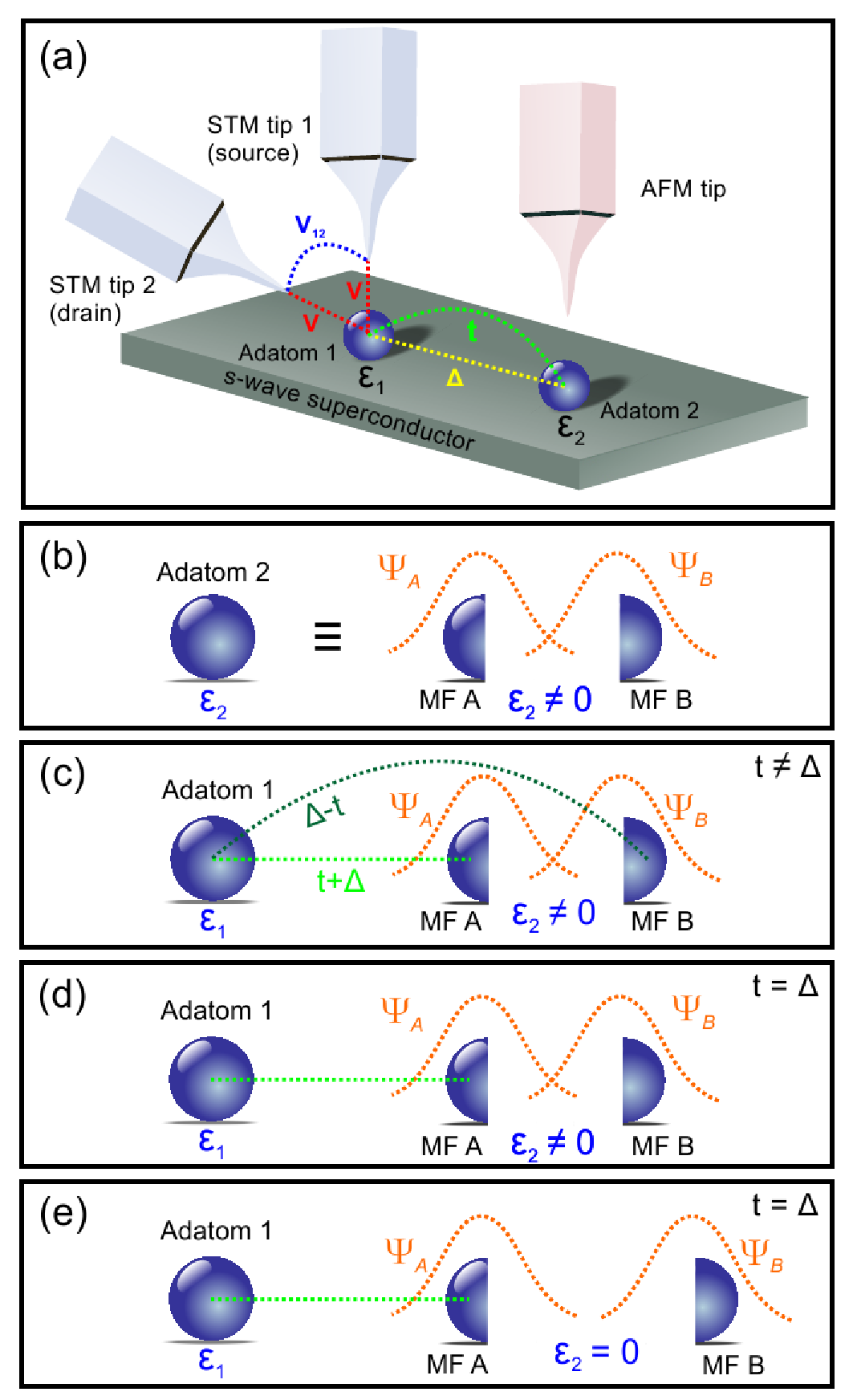} \protect\protect\caption{\label{Fig1}(Color online) (a) Setup composed by STM and AFM tips
in the presence of superconducting adatoms hosted by a conventional
superconductor with strong spin-orbit coupling. (b) The adatom 2 is
made by a pair of coupled Majorana fermions (MFs $A$ and $B$ represented
by half-spheres) where the level $\varepsilon_{2}$ is induced by
the AFM tip and plays the role of the connection between the MFs.
It can assume negative values when it stays below the MF zero mode,
due to the tuning performed by the gate potential of the AFM tip.
(c) In the case of $t\protect\neq\Delta$ and $\varepsilon_{2}\protect\neq0,$
both MFs hybridize with the adatom 1. (d) For the matching of the
Cooper pair binding energy with the normal tunneling between the adatoms
1 and 2 $(t=\Delta)$ and $\varepsilon_{2}\protect\neq0,$ the MF
$A$ becomes the unique Majorana connected to the adatom 1. (e) Here
we see the isolated MF $A$ coupled to the adatom 1 when $t=\Delta$
and $\varepsilon_{2}=0,$ which results in the standard ZBP in the
transmittance and in the isolated MF $B$ at one edge of the Kitaev
dimer.}

\par\end{centering}

\centering{}
\end{figure}

Experimentally, $p$-wave superconductivity is feasible due to the
proximity effect by the employment of an $s$-wave superconductor
close to a semiconducting nanowire characterized by a spin-orbit interaction
under an external magnetic field \citep{key-DLoss1,key-DLoss2,key-PSodano1,key-Nayana}.
Particularly in the case of transport through quantum dots (QDs) coupled
to a MF \citep{key-371,Zheng,JAP1,JAP2,key-25,Liu,Cheng,key-25b,key-25c,Jap-2014,Ueda,Zhang,key-RA2},
a zero-bias peak (ZBP) \citep{key-25,key-25b} in the conductance is
expected to be observed. It is worth mentioning that the ZBP has been
detected in conductance measurements through a nanowire of indium
antimonide linked to gold and niobium titanium nitride electrodes
\citep{key-301}. Analogously, a ZBP has also been verified in the
superconducting system of aluminium next to a nanowire of indium arsenide
\citep{key-302}. However, the ZBP signature may also have another
physical origin, for instance the Kondo effect \citep{STM12,STM5,STM6,AHCNeto,STM13},
thus turning the experimental ZBP detection inconclusive within a
MF perspective. Moreover, recently an alternative way for the achievement
of the topological Kitaev chain has been the employment of magnetic
chains on top of superconductors \citep{c1,c2,c3,c4,Glazman,Jelena2}.
Particularly in Ref.\citep{c4}, the ZBP observed
exhibits a subtle amplitude of the order $10^{-4}(2G_{0})$, which
is a signal extremely weak due to thermal broadening, where $G_{0}=e^{2}/h$
is the quantum of conductance. Thus in the current context, novel
approaches in the pursuit of MFs become necessary. In this scenario,
we highlight the proposal found in Ref.\citep{Nano},
which is a pioneering example concerning this issue: a hybrid spin-microcantilever
system for optical detection of MFs on the edges of a Kitaev chain.

In this work we propose a new route for detection of MFs signatures.
To that end, we consider the setup outlined in Fig.\,\ref{Fig1}
where an \textit{{s}}-wave superconductor
with strong spin-orbit coupling hosts a set of magnetic adatoms, in
analogy to the experimental apparatus developed in Ref.\citep{c4}
that describes a superconducting surface of lead (Pb) well known by
its strong spin-orbit coupling, thus allowing \textit{p-}wave
superconductivity on this chain due to the proximity effect. It is
worth mentioning that the proposal of Ref.\citep{c4}
is distinct from the semiconducting spin-orbit wire reported in Ref.\citep{{key-301}}.
In the current work, we consider for a sake of simplicity, just a
pair of superconducting adatoms (Kitaev dimer), in particular placed
nearby STM (Scanning Tunneling Microscope) and AFM (Atomic Force Microscope)
tips. Additionally, we assume two highly spin-polarized STM tips
in order to freeze the spin degree of freedom, thus avoiding the emergence
of the Kondo effect within the adatom 1.

In regard to the possibility of experimental realization of the setup
shown in Fig.\,\ref{Fig1}, we stress that multitip STM experiments
can be possible, see e.g. Refs.\citep{{m1,m2,m3}}.
In the case of the AFM tip, which is fixed on top of an adatom, it
operates similarly to the Scanning Gate Microscopy (SGM) technique
\citep{SGM}, wherein a charged tip allows that its gate potential
tunes the energy levels of the adatom probed. This approach is widely
employed in systems with quantum point contacts (QPCs) formed in two
dimensional electron gases \citep{QPC,QPC1}, which we invoke here
to our proposal. Thereby despite the challenging of applying the SGM
technique in the scenario of Fig.\,\ref{Fig1}, we trust that in
the near future such a procedure can be implemented.

In the frame of the setup here proposed, a device based on Fano interference
\citep{Fano1,Fano2,Ueda,Zhang} becomes an alternative method for detection
of MFs. As a matter of fact, the Fano effect is found in systems where
distinct tunneling paths compete for the electron transport. Noteworthy,
such a phenomenon can be perceived experimentally by the STM just
by measuring the conductance \citep{STM12,STM5,STM6,AHCNeto,STM13}.
In the setup of Fig.\ref{Fig1} here proposed, the
Fano effect arises from the adatom 1 coupled simultaneously to the
pair of STM tips, once the electrons can tunnel between such tips
or (and) directly through this adatom as expected for realistic experimental
conditions. Thus depending on the Fano lineshape of the conductance,
we can identify which path rules the quantum transport or if they
compete on the same footing. Later on, we will see that the Fano parameter
of interference $q_{b}$ then dictates the Fano profile making explicit
the dominant path in the system.

Here we show that the two coupled MFs $A$ and $B$ within the adatom
under the AFM tip (see Fig.\ref{Fig1}) lead to a gate invariance
feature in the transport experiment. Particularly,
when the AFM tip swaps symmetrically its gate potential around the
MF zero mode, it reveals a universality signature in the Fano profile
of the transmittance. More specifically, we find then distinct situations
in which the transmittance shares the same profile as a function of
the Fermi level of the STM tips, in particular when the binding energy
of the Cooper pair split into the adatoms under the STM and AFM tips,
is in resonance with the tunneling term between them. In this case,
we show that when the MF $A$ couples solely to the adatom $1$, the
gate invariance characterized by the aforementioned universality is
revealed by connecting MFs $A$ and $B$.

This paper is organized as follows: in Sec.\,\ref{sec2} we develop
the theoretical model for the system sketched in Fig.\,\ref{Fig1}
by deriving the expression for the transmittance through such a device
(see also the Appendix) together with the system Green's functions.
The results are present in Sec.\,\ref{sec3}, and in Sec.\,\ref{sec4}
we summarize our concluding remarks.

\section{The model}

\label{sec2}

\subsection{System Hamiltonian}

Here we consider the system outlined in Fig.\,\ref{Fig1}(a) for
a pair of highly spin-polarized STM tips connected to the hybrid setup
composed by an AFM tip and the ``host+superconducting adatoms''
set. Thus such a spinless model reads:
\begin{align}
\mathcal{H} & =  \underset{\alpha k}{\sum}\tilde{\varepsilon}_{\alpha k}c_{\alpha k}^{\dagger}c_{\alpha k}+\sum_{j}\varepsilon_{j}d_{j}^{\dagger}d_{j}\nonumber \\
& +  V\underset{\alpha k}{\sum}(c_{\alpha k}^{\dagger}d_{1}+\text{{H.c.}})\nonumber \\
 & +  (V_{12}\sum_{kq}c_{1k}^{\dagger}c_{2q}+td_{1}d_{2}^{\dagger}+\Delta d_{2}^{\dagger}d_{1}^{\dagger}+\text{{H.c.}}),\label{eq:TIAM}
\end{align}
where the electrons in the STM tip $\alpha=1,2$ (STM tip $1$ and
$2$, respectively) are described by the operator $c_{\alpha k}^{\dagger}$
($c_{\alpha k}$) for the creation (annihilation) of an electron in
a quantum state labeled by the wave number $k$ and energy $\tilde{\varepsilon}_{\alpha k}=\varepsilon_{k}-\mu_{\alpha}$,
with $\mu_{\alpha}$ as the chemical potential. Here we adopt the
gauge $\mu_{2}=\Delta\mu$ and $\mu_{1}=-\Delta\mu$, with $\mu_{2}-\mu_{1}=2\Delta\mu=e\varphi$
as the bias between the tips, being $e>0$ the electron charge and
$\varphi$ the bias-voltage. Consequently, the transmittance through
the setup is a function of the Fermi energy $\varepsilon=\mu_{1}=\mu_{2}$
of the STM tips, where the point $\varepsilon=0$ corresponds to the
MF zero mode. For the adatoms, $d_{j}^{\dagger}$ ($d_{j}$) creates
(annihilates) an electron in the state $\varepsilon_{j}$, with $j=1,2$.
$V$ stands for the hybridizations between the adatom 1 and the STM
tips. These couplings are considered the same to avoid Andreev currents
in the system \citep{key-371,Zheng} (see also the Appendix). $V_{12}$
is the STM tip 1-tip 2 coupling, which will ensure the precise renormalization
of the energy level in the adatom $1$ as we will see.

The \textit{s-}wave superconductor with strong spin-orbit coupling
enters into the model via the parameters $t$ and $\Delta,$ which
respectively yield the hopping term between the adatoms 1 and 2, and
also the binding energy of a delocalized Cooper pair split into such
adatoms: as a net effect of the arrangement of Fig.\,\ref{Fig1}(a)
considered, \textit{p-}wave superconductivity is induced on the pair
of adatoms (Kitaev dimer) close to the tips, similarly to that reported
in Ref.\citep{{c4}} for the case of a long Kitaev chain.
As the adatom 1 is placed within the region separating the STM tips
and the superconducting surface, we can safely neglect the tunneling
from the former into the latter as an outcome of the setup adopted.
We stress that the role of the AFM tip is indeed to gate overlap the
MFs enclosed by the adatom 2. This \textit{clue} between the MFs found
within such an adatom will be clarified later on as the own energy
level of this adatom, which consists in a degree of freedom completely
tunable by the AFM tip. Thus for a sake of simplicity, we clamped
these features as the most relevant from the proximity problem between
the \textit{s-}wave surface and the adatoms. Alternatively to this
end, an approach concerning \textit{ab-initio} description on the
proximity issue can be helpful, but it does not belong to the current
scope of the paper.

Here we express the adatom 2 in the Majorana basis by following the
transformation $d_{2}=\frac{1}{\sqrt{2}}(\Psi_{A}+i\Psi_{B})$ and
$d_{2}^{\dagger}=\frac{1}{\sqrt{2}}(\Psi_{A}-i\Psi_{B}),$ in which
$\Psi_{l}^{\dagger}=\Psi_{l}$ $(l=A,B)$ characterizes a MF operator,
thus yielding
\begin{align}
\mathcal{H}_{\text{{MFs}}} & =\varepsilon_{2}d_{2}^{\dagger}d_{2}+(td_{1}d_{2}^{\dagger}+\Delta d_{2}^{\dagger}d_{1}^{\dagger}+\text{{H.c.}})\nonumber \\
 & =i\varepsilon_{2}\Psi_{A}\Psi_{B}+\frac{(t+\Delta)}{\sqrt{2}}(d_{1}-d_{1}^{\dagger})\Psi_{A}\nonumber \\
 & +i\frac{(\Delta-t)}{\sqrt{2}}(d_{1}+d_{1}^{\dagger})\Psi_{B}+\frac{\varepsilon_{2}}{2},\label{eq:MQPs}
\end{align}
where the first term in the second line makes explicit the MFs within
the adatom $2$ and the others stand for the connections between MFs
$A$ and $B$ with the adatom $1$ (see Figs.\ref{Fig1}(c), (d) and
(e)). It is worth noticing that the standard Majorana
ZBP occurs for $t=\Delta,$ $\varepsilon_{1}\neq0$ and $\varepsilon_{2}=0,$
with the conductance $G=0.5G_{0}$ due to the MF $\gamma_{1}=i(d_{1}-d_{1}^{\dagger})/\sqrt{2}$
existing in the adatom 1 given by the second term of Eq.(\ref{eq:MQPs})
($-2it\gamma_{1}\Psi_{A}$), which has a component on the adatom 2,
where a completely localized MF is observed according to M. Leijnse
and K. Flensberg in Ref.\citep{key-377}.

In this scenario, we explore the model regimes $t=\Delta$ and $t\neq\Delta$
combined with $\varepsilon_{2}>0$ and $\varepsilon_{2}<0$ varied
symmetrically with respect to the MF zero mode. We stress that for
a fixed $\Delta,$ the tuning of the parameter $t$ can be performed
experimentally just by changing the distance between the adatoms 1
and 2, while the AFM tip controls the level position $\varepsilon_{2}$
of the adatom 2, since such a tip operates as a gate potential. Particularly
for $t=\Delta,$ we can verify in Eq.\,(\ref{eq:MQPs}) that the
adatom 1 decouples from the MF $B$ (here sketched by the most right
half-sphere in Fig.\,\ref{Fig1}(d)) thus allowing an exclusively
connection between this adatom and the MF $A$ (the most left half-sphere)
as the Hamiltonian $\mathcal{H}_{\text{{MFs}}}=\varepsilon_{2}(i\Psi_{A}\Psi_{B}+\frac{1}{2})+\sqrt{2}\Delta(d_{1}-d_{1}^{\dagger})\Psi_{A}$
points out. For $\varepsilon_{2}\neq0,$ the Fano profile of the transmittance
becomes invariant under the symmetric change of $\varepsilon_{2}$
with respect to the MF zero mode, in contrast with the case $t\neq\Delta$
where this universality is prevented. Thereafter, it gives rise to
distinct transmittance profiles strongly dependent on the sign of
$\varepsilon_{2}$ away from the point $t=\Delta.$

Thereby we recognize this invariance as a Majorana property of the
system when it is driven to the point $t=\Delta$ with symmetric swap
of $\varepsilon_{2}.$

\subsection{Conductance}

In what follows we derive the Landauer-Büttiker formula for the zero-bias
conductance $G$ \citep{book}. Such a quantity is a function of the
transmittance $\mathcal{T}\left(\varepsilon\right)$ as follows:
\begin{equation}
G=G_{0}\int d\varepsilon\left(-\frac{\partial f_{F}}{\partial\varepsilon}\right)\mathcal{T}(\varepsilon),\label{eq:_10b}
\end{equation}
where $f_{F}$ stands for the Fermi-Dirac distribution.

We begin with the transformations $c_{2k}=\frac{1}{\sqrt{2}}(c_{ek}+c_{ok})$
and $c_{1k}=\frac{1}{\sqrt{2}}(c_{ek}-c_{ok})$ on the Hamiltonian
of Eq.\,(\ref{eq:TIAM}), which starts to depend on the \textit{even}
and \textit{odd} conduction operators $c_{ek}$ and $c_{ok}$, respectively.
These definitions allow us to express Eq.\,(\ref{eq:TIAM}) as $\mathcal{H}=\mathcal{H}_{e}+\mathcal{H}_{o}+\mathcal{\tilde{H}}_{\text{{tun}}}=\mathcal{H}_{\varphi=0}+\mathcal{\tilde{H}}_{\text{{tun}}},$
where
\begin{align}
\mathcal{H}_{e} & = \sum_{k}\varepsilon_{k}c_{ek}^{\dagger}c_{ek}+\varepsilon_{1}d_{1}^{\dagger}d_{1} + V_{12}\sum_{kp}c_{ek}^{\dagger}c_{ep}\nonumber \\
 & +\sqrt{2}V\sum_{k}(c_{ek}^{\dagger}d_{1}+\text{{H.c.}})+\mathcal{H}_{\text{{MFs}}}\label{eq:even}
\end{align}
represents the Hamiltonian part of the system coupled to the adatoms
via an effective hybridization $\sqrt{2}V$, while
\begin{equation}
\mathcal{H}_{o}=\sum_{k}\varepsilon_{k}c_{ok}^{\dagger}c_{ok}-V_{12}\sum_{kp}c_{ok}^{\dagger}c_{op}\label{eq:odd}
\end{equation}
is the decoupled one. However, they are connected to each other by
the tunneling Hamiltonian $\mathcal{\tilde{H}}_{\text{{tun}}}=-\Delta\mu\sum_{k}(c_{ek}^{\dagger}c_{ok}+c_{ok}^{\dagger}c_{ek}).$

As in the zero-bias regime $\Delta\mu\rightarrow0$, due to $\varphi\rightarrow0$,
$\mathcal{\tilde{H}}_{\text{{tun}}}$ is a perturbative term and the
linear response theory (see the Appendix) ensures that
\begin{equation}
\mathcal{T}\left(\varepsilon\right)=(2\pi V_{12})^{2}\tilde{\rho}_{e}(\varepsilon)\tilde{\rho}_{o}(\varepsilon),\label{eq:transmit-1}
\end{equation}
where $\tilde{\rho}_{e}(\varepsilon)=-\frac{1}{\pi}{\tt Im}(\tilde{\mathcal{G}}_{\Psi_{e}\Psi_{e}})$
is the local density of states (LDOS) for the Hamiltonian of Eq.\,(\ref{eq:even})
and
\begin{align}
\mathcal{G}_{\Psi_{e}\Psi_{e}} & =-\frac{i}{\hbar}\theta\left(\tau\right){\tt Tr}\{\varrho_{\text{e}}[\Psi_{e}\left(\tau\right),\Psi_{e}^{\dagger}\left(0\right)]_{+}\}\label{eq:PSI_R-2}
\end{align}
gives the retarded Green's function in the time domain $\tau$, where
$\theta(\tau)$ is the Heaviside step function, $\varrho_{\text{e}}$
is the density-matrix for Eq.\,($\ref{eq:even}$), $\Psi_{e}=f_{e}+(\pi\Gamma\rho_{0})^{1/2}qd_{1}$
is a field operator, with $f_{e}=\sum_{p}c_{ep},$ the Anderson parameter
$\Gamma=2\pi V^{2}\rho_{0},$ with $\rho_{0}$ as the density of states
for the STM tips and $q=(\pi\rho_{0}\Gamma)^{-1/2}\left(\frac{\sqrt{2}V}{2V_{12}}\right).$

To calculate Eq.\,(\ref{eq:PSI_R-2}) in the energy domain $\varepsilon$,
we should employ the equation-of-motion (EOM) method \citep{book}
summarized as follows
\begin{equation}
(\varepsilon+i0^{+})\tilde{\mathcal{G}}_{\mathcal{AB}}=[\mathcal{A},\mathcal{B^{\dagger}}]_{+}+\tilde{\mathcal{G}}_{\left[\mathcal{A},\mathcal{\mathcal{H}}_{i}\right]\mathcal{B}}\label{eq:EOM}
\end{equation}
for the retarded Green's function $\tilde{\mathcal{G}}_{\mathcal{AB}}=\int d\tau\mathcal{G}_{\mathcal{AB}}e^{\frac{i}{\hbar}(\varepsilon+i0^{+})\tau},$
which is the time Fourier transform of $\mathcal{G}_{\mathcal{AB}},$
with $\mathcal{A}$ and $\mathcal{B}$ as fermionic operators belonging
to the Hamiltonian $\mathcal{\mathcal{H}}_{i}.$ By considering $\mathcal{A=B}=\Psi{}_{e}$
and $\mathcal{H}_{i}=\mathcal{H}_{e}$, we find
\begin{align}
\tilde{\mathcal{G}}_{\Psi_{e}\Psi_{e}} & =\tilde{\mathcal{G}}_{f_{e}f_{e}}+(\pi\rho_{0}\Gamma)q^{2}\tilde{\mathcal{G}}_{d_{1}d_{1}}+2(\pi\rho_{0}\Gamma)^{1/2}q\tilde{\mathcal{G}}_{d_{1}f_{e}}.\nonumber \\
\label{eq:25b}
\end{align}
From Eqs.\,(\ref{eq:even}), (\ref{eq:EOM}) with $\mathcal{A=B}=f_{e}$
and (\ref{eq:25b}), we obtain
\begin{align}
\tilde{\mathcal{G}}_{f_{e}f{}_{e}}&= \frac{\pi\rho_{0}(\bar{q}-i)}{1-\sqrt{x}(\bar{q}-i)}\nonumber \\
& +  \pi\rho_{0}\Gamma\left[\frac{(\bar{q}-i)}{1-\sqrt{x}(\bar{q}-i)}\right]^{2}\tilde{\mathcal{G}}_{d_{1}d_{1}}\label{eq:g_ff}
\end{align}
and the mixed Green's function
\begin{equation}
\tilde{\mathcal{G}}_{d_{1}f_{e}}=\sqrt{\pi\Gamma\rho_{0}}\frac{(\bar{q}-i)}{1-\sqrt{x}(\bar{q}-i)}\tilde{\mathcal{G}}_{d_{1}d_{1}},\label{eq:g_df}
\end{equation}
determined from Eq.\,(\ref{eq:EOM}) by considering $\mathcal{A}=d_{1}$,
$\mathcal{B}=f{}_{e}$ and $\mathcal{H}_{i}=\mathcal{H}_{e}$, with
the parameter $x=(\pi\rho_{0}V_{12})^{2}$ and $\bar{q}=\frac{1}{\pi\rho_{0}}\sum_{k}\frac{1}{\varepsilon-\varepsilon_{k}}.$
Here we assume the wide band limit denoted by $\bar{q}\rightarrow0.$

Additionally, for the Hamiltonian of Eq.\,(\ref{eq:odd}) we have
the LDOS $\tilde{\rho}_{o}(\varepsilon)=-\frac{1}{\pi}{\tt Im}(\tilde{\mathcal{G}}_{f_{o}f_{o}}),$
with
\begin{align}
\mathcal{G}_{f_{o}f_{o}} & =-\frac{i}{\hbar}\theta\left(\tau\right){\tt Tr}\{\varrho_{\text{o}}[f_{o}\left(\tau\right),f_{o}^{\dagger}\left(0\right)]_{+}\}\label{eq:PSI_R-2-1}
\end{align}
and $f_{o}=\sum_{\tilde{q}}c_{o\tilde{q}}.$ We notice that $\tilde{\mathcal{G}}_{f_{o}f_{o}}$
is decoupled from the adatoms. Thereby, from Eqs.\,(\ref{eq:odd})
and (\ref{eq:PSI_R-2-1}), we take $\mathcal{A}=\mathcal{B}=f{}_{o}$
and $\mathcal{H}_{i}=\mathcal{H}_{o}$ in Eq.\,(\ref{eq:EOM}) and
we obtain
\begin{equation}
\tilde{\mathcal{G}}_{f_{o}f{}_{o}}=\frac{\pi\rho_{0}(\bar{q}-i)}{1+\sqrt{x}(\bar{q}-i)}.\label{eq:rho_ff}
\end{equation}
Thus the substitution of Eqs.\,(\ref{eq:25b}), (\ref{eq:g_df}),
and (\ref{eq:rho_ff}) in Eq.\,(\ref{eq:transmit-1}), leads to
\begin{align}
\frac{\mathcal{T}\left(\varepsilon\right)}{\mathcal{T}_{b}} & =1+(1-q_{b}^{2})\tilde{\Gamma}\text{{Im}}(\tilde{\mathcal{G}}_{d_{1},d_{1}})+2q_{b}\tilde{\Gamma}\text{{Re}}(\tilde{\mathcal{G}}_{d_{1},d_{1}}),\label{eq:trans2}
\end{align}
where $\tilde{\Gamma}=\frac{\Gamma}{1+x}$ is an effective adatom
1-tip coupling, $\mathcal{T}_{b}=\frac{4x}{\left(1+x\right)^{2}}$
represents the transmittance through the STM tips when the adatom
1 is absent ($\tilde{\Gamma}=0$), $\mathcal{R}_{b}=1-\mathcal{T}_{b}$
stands for the corresponding reflectance and $q_{b}=\sqrt{\frac{\mathcal{R}_{b}}{\mathcal{T}_{b}}}=\frac{\left(1-x\right)}{2\sqrt{x}}$
as the Fano parameter \citep{Fano1,Fano2,key-26}. The current equation
for the transmittance encodes three distinct Fano regimes of interference
as follows: i) $q_{b}\rightarrow\infty$ ($x=0$) provides $\mathcal{T}\left(\varepsilon\right)=-\tilde{\Gamma}\text{{Im}}(\tilde{\mathcal{G}}_{d_{1},d_{1}})=\pi\tilde{\Gamma}\rho_{11},$
where we use $\rho_{11}=-\frac{1}{\pi}{\tt Im}(\tilde{\mathcal{G}}_{d_{1},d_{1}})$
as the LDOS of the adatom 1, which contains resonant states characterized
by peaks, since the electronic transport in the interferometer is
solely through the adatom 1 \citep{key-25}; ii) $q_{b}=0$ ($x=1$)
leads to $\mathcal{T}\left(\varepsilon\right)=1-\pi\tilde{\Gamma}\rho_{11}$
exhibiting Fano dips as a result of the suppression caused by $\rho_{11}$
over the first term representing the maximum amplitude $\mathcal{T}_{b}=1.$
In such a case, when $\rho_{11}$ shows a maximum, $\mathcal{T}\left(\varepsilon\right)$
presents a corresponding minimum as expected and the tunneling between
the STM tips becomes the dominant process in the system. It reveals
the depletion of charge in the LDOS of the STM tips detected by the
transmittance, once this charge accumulates within the adatom 1, in
particular, around the resonant states of $\rho_{11}$ observed in
situation (i) for $q_{b}\rightarrow\infty$ ($x=0$); iii) corresponds
to $q_{b}\approx0.35$ ($x=0.5$), which is the intermediate case
wherein asymmetric Fano lineshapes appear with peaks and Fano dips
coexisting in the same profile, thus making explicit a competition
between $V$ and $V_{12}$ on an equal footing. These features of
Fano interference will be addressed in Sec.\ref{sec3}.

\subsection{System Green's functions}

By applying the EOM on
\begin{align}
\mathcal{G}_{d_{1}d_{1}} & =-\frac{i}{\hbar}\theta\left(\tau\right){\tt Tr}\{\varrho_{\text{{e}}}[d_{1}\left(\tau\right),d_{1}^{\dagger}\left(0\right)]_{+}\},\label{eq:djdl}
\end{align}
and changing to the energy domain $\varepsilon$, we obtain the following
relation
\begin{align}
(\varepsilon-\varepsilon_{1}-\Sigma)\tilde{\mathcal{G}}_{d_{1}d_{1}} & =1-t\mathcal{\tilde{G}}_{d_{2},d_{1}}-\Delta\mathcal{\tilde{G}}_{d_{2}^{\dagger},d_{1}}\label{eq:GF11}
\end{align}
expressed in terms of the self-energy $\Sigma=-(\sqrt{x}+i)\tilde{\Gamma}$
and Green's functions $\mathcal{\tilde{G}}_{d_{2},d_{1}}$ and $\mathcal{\tilde{G}}_{d_{2}^{\dagger},d_{1}}.$
According to the EOM approach we find

\begin{eqnarray}
\mathcal{\tilde{G}}_{d_{2},d_{1}} & = &  -\frac{t\mathcal{\tilde{G}}_{d_{1},d_{1}}}{(\varepsilon-\varepsilon_{2}+i0^{+})}+\frac{\Delta\mathcal{\tilde{G}}_{d_{1}^{\dagger},d_{1}}}{(\varepsilon-\varepsilon_{2}+i0^{+})},\nonumber \\
&  & \label{eq:GF21}\\
\mathcal{\tilde{G}}_{d_{2}^{\dagger},d_{1}} & = &  -\frac{\Delta\mathcal{\tilde{G}}_{d_{1},d_{1}}}{(\varepsilon+\varepsilon_{2}+i0^{+})}+\frac{t\mathcal{\tilde{G}}_{d_{1}^{\dagger},d_{1}}}{(\varepsilon+\varepsilon_{2}+i0^{+})}\nonumber \\
 &  &\label{eq:GF21plus}
\end{eqnarray}
and

\begin{equation}
\mathcal{\tilde{G}}_{d_{1}^{\dagger},d_{1}}=-2t\Delta\tilde{K}\mathcal{\tilde{G}}_{d_{1},d_{1}},\label{eq:GF11plus}
\end{equation}
in which $\tilde{K}=\frac{K}{\varepsilon+\varepsilon_{1}+\bar{\Sigma}-K_{-}},$
with $K=\frac{(\varepsilon+i0^{+})}{[\varepsilon^{2}-\varepsilon_{2}^{2}+2i\varepsilon0^{+}-(0^{+})^{2}]},$
$\bar{\Sigma}$ as the complex conjugate of $\Sigma$ and $K_{\pm}=\frac{(\varepsilon+i0^{+})(t^{2}+\Delta^{2})\pm\varepsilon_{2}(t^{2}-\Delta^{2})}{[\varepsilon^{2}-\varepsilon_{2}^{2}+2i\varepsilon0^{+}-(0^{+})^{2}]}.$
Thus substituting Eqs.\,(\ref{eq:GF21}), (\ref{eq:GF21plus}) and
(\ref{eq:GF11plus}) into Eq. (\ref{eq:GF11}) the Green's function
of the adatom $1$ becomes

\begin{align}
\tilde{\mathcal{G}}_{d_{1}d_{1}} & =\frac{1}{\varepsilon-\varepsilon_{1}-\Sigma-\Sigma_{\text{{MFs}}}},\label{eq:d1d1}
\end{align}
where

\begin{equation}
\Sigma_{\text{{MFs}}}=K_{+}+(2t\Delta)^{2}K\tilde{K}\label{eq:MQPself}
\end{equation}
accounts for the self-energy due to the MFs connected to the adatom
1. We highlight that the self-energy of Eq. (\ref{eq:MQPself}) contains
the underlying mechanism that allows the invariance of the Fano lineshape.
Such a universal feature is revealed as independent on the Fano parameter
$q_{b},$ which we will discuss in detail below. Particularly for
$t=\Delta=\frac{\lambda}{\sqrt{2}},$ we highlight that the expressions
for $\tilde{K}$ and $\Sigma_{\text{{MFs}}}$ found in Ref.\citep{key-25} are recovered. Such a result will be revisited in Sec.\ref{sec3}.

\section{Results and discussion}

\label{sec3}

Below we investigate the features of the system Green's functions
by employing the expression for the transmittance (Eq.\,(\ref{eq:trans2})).
According to Eq.\,(\ref{eq:_10b}), this transmittance can be obtained
experimentally via the conductance $G$ in units of $G_{0}$ for temperatures
$T\rightarrow0.$ Additionally, we employ values for the Fermi energy
$\varepsilon,$ $\varepsilon_{j},$ $t,$ and $\Delta$ in units of
the Anderson parameter $\Gamma.$

\begin{figure}[!htb]
\includegraphics[width=0.5\textwidth,height=0.3\textheight]{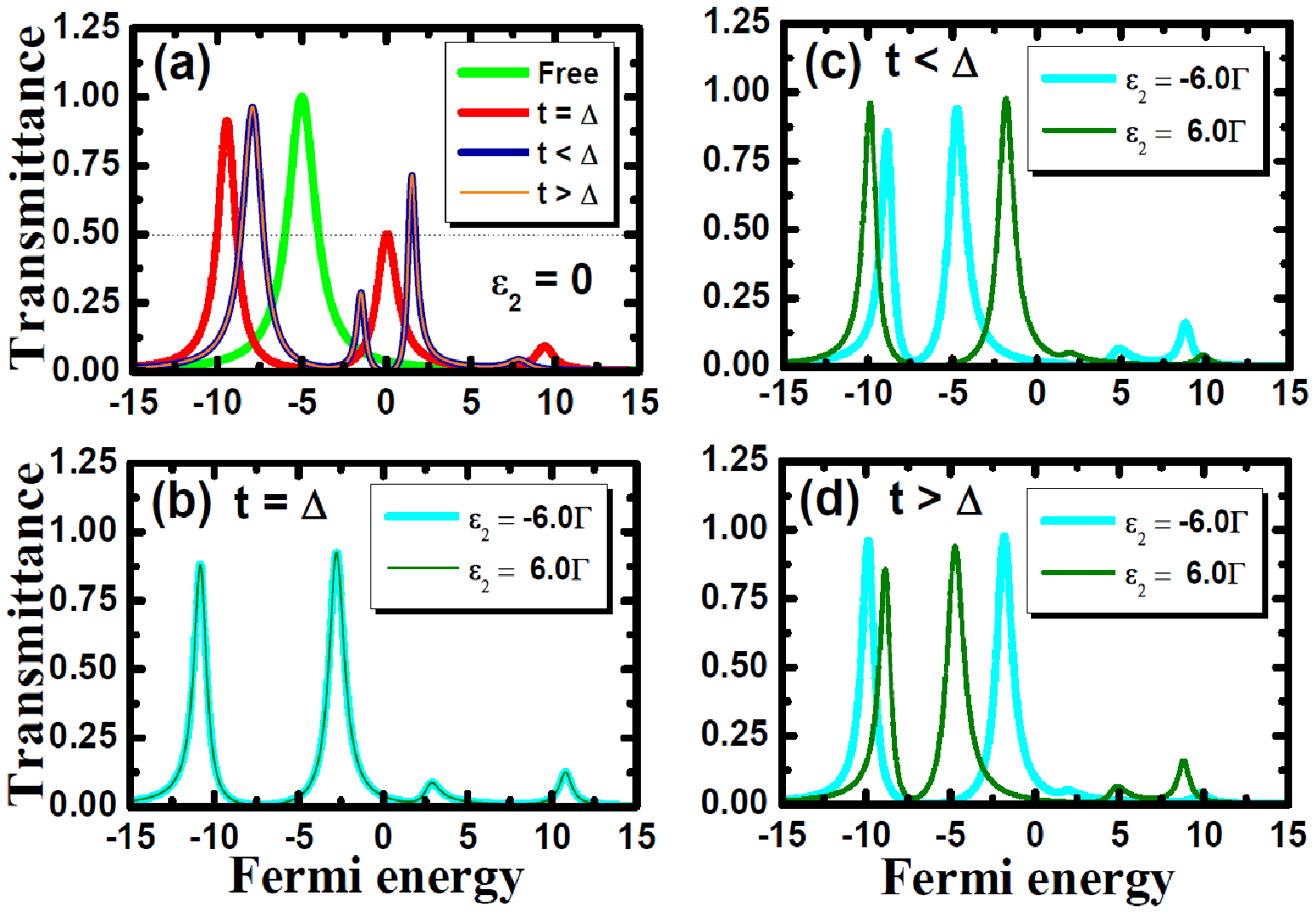}
\protect\protect\protect\protect\protect\protect\protect\protect\protect\protect\protect\protect\protect\protect\protect\protect\protect\protect\protect\protect\protect\protect\protect\protect\protect\protect\protect\protect\protect\protect\protect\caption{\label{Fig2}(Color online) Transmittance as a function of the Fermi
energy of the STM tips within the Fano regime $x=0$ $(q_{b}\rightarrow\infty)$:
(a) for several cases in the parameters $t$ and $\Delta$ with $\varepsilon_{2}=0.$
Particularly in the condition $t=\Delta,$ the standard zero-bias
peak is found. (b) For $t=\Delta$ the transmittance shares the same
lineshape of the symmetric situations above $(\varepsilon_{2}=6\Gamma)$
and below $(\varepsilon_{2}=-6\Gamma)$ the MF zero mode $\varepsilon=0$,
respectively for positive and negative AFM gate potentials. The transmittance
then becomes an invariant quantity under these conditions. Panels
(c) and (d) reveal distinct profiles when the system is driven away
$(t\protect\neq\Delta)$ from the point $t=\Delta$.}
\end{figure}

In Fig.\ref{Fig2} we consider the Fano regime $x=0$ $(q_{b}\rightarrow\infty)$
for the transmittance $\mathcal{T}$ of Eq.\,(\ref{eq:trans2}) as
a function of the Fermi energy $\varepsilon.$ This situation corresponds
to the case where the electron tunneling occurs exclusively through
the adatom 1, due to the strong coupling between it and the STM tips.
As predicted by the standard Fano's theory \citep{Fano1}, the transmittance
should exhibit a peak around each localized state in the adatom probed
by the tips: see the green line shape of panel (a) for the adatom
1 here assumed to be decoupled from the adatom 2 for a sake of simplicity,
which leads to the resonance centered at $\varepsilon=\varepsilon_{1}=-5$
with maximum amplitude $\mathcal{T}=1.$ By keeping
this level at such a value and employing $t=\Delta=4$ combined with
$\varepsilon_{2}=0,$ a ZBP given by $\mathcal{T}=1/2$ emerges due
to the MF existing in adatom 1\citep{key-377}. Additionally, the
most left resonance in the same curve corresponds to that at $\varepsilon=-5$
found in the green lineshape, in particular with renormalized peak
position $\varepsilon\approx-10$ as the aftermath of the connection
$\sqrt{2}\Delta$ with the adatom 2, and with higher amplitude $(\mathcal{T}>1/2)$
in respect to that for the ZBP $(\mathcal{T}=1/2).$ Notice that a
third peak in the vicinity of $\varepsilon\approx+10$ is found characterized
by $\mathcal{T}<1/2.$ Thus in the presence of finite couplings $t,$
the original peak at $\varepsilon=\varepsilon_{1}=-5$ in the green
curve of panel (a) with $\mathcal{T}=1$ is split into those at $\varepsilon\approx-10$
and $\varepsilon\approx+10$ both with $\mathcal{T}<1$ as the red
lineshape points out. In which concerns the curves for $t<\Delta$
($t=2$ and $\Delta=4$) and $t>\Delta$ ($t=4$ and $\Delta=2$),
the transmittance is revealed as independent on the strengths $t$
and $\Delta.$ Such a behavior attests the situation in which coupled
MFs are absent in the system within the adatom 2.

Panel (b) of Fig.\ref{Fig2} depicts the situation in which the system
is still within the regime $t=\Delta,$ but with $\varepsilon_{2}\neq0.$
We point out that such a panel reveals a universal
behavior in the transmittance profile when symmetric values for $\varepsilon_{2}$
are accounted. This regime is characterized by coupled MFs (see Fig.\,\ref{Fig1}(d))
which result in the suppression of the ZBP and the splitting of the
resonances at $\varepsilon\approx-10$ and $\varepsilon\approx+10$
observed in the red curve of panel (a). For instance, the aforementioned
universality is verified providing two identical curves for both values
$\varepsilon_{2}=6\Gamma$ (positive potential) and $\varepsilon_{2}=-6\Gamma$
(negative potential) due to the self-energy $\Sigma_{\text{{MFs}}}$
of Eq.\,(\ref{eq:MQPself}) for the MFs, which is dependent on the
amplitude $K_{\pm}$. Notice that $K_{\pm}=\frac{2t^{2}(\varepsilon+i0^{+})}{[\varepsilon^{2}-\varepsilon_{2}^{2}+2i\varepsilon0^{+}-(0^{+})^{2}]}$
within this situation, thus implying in $K_{\pm}(\varepsilon_{2})=K_{\pm}(-\varepsilon_{2})$
as well as $\Sigma_{\text{{MFs}}}(\varepsilon_{2})=\Sigma_{\text{{MFs}}}(-\varepsilon_{2}),$
which then ensure the invariance of the transmittance profile at the
point $t=\Delta.$ For $t\neq\Delta$ we have $K_{\pm}\propto\pm\varepsilon_{2}(t^{2}-\Delta^{2}),$
thus allowing a strong dependence on the sign of $\varepsilon_{2}.$
In panels (c) and (d) of the same figure, the transmittance respectively
for $t<\Delta$ ($t=2$ and $\Delta=4$) and $t>\Delta$ ($t=4$ and
$\Delta=2$) exhibit distinct behaviors as expected when we adopt
symmetric values $\varepsilon_{2}=6\Gamma$ and $\varepsilon_{2}=-6\Gamma.$
Therefore in both limits $t>\Delta$ and $t<\Delta,$ the influence
of the negative potential $\varepsilon_{2}<0$ on the transmittance
is made explicit once $\Sigma_{\text{{MFs}}}(\varepsilon_{2})\neq\Sigma_{\text{{MFs}}}(-\varepsilon_{2}).$

\begin{figure}[!htb]
\includegraphics[width=0.5\textwidth,height=0.3\textheight]{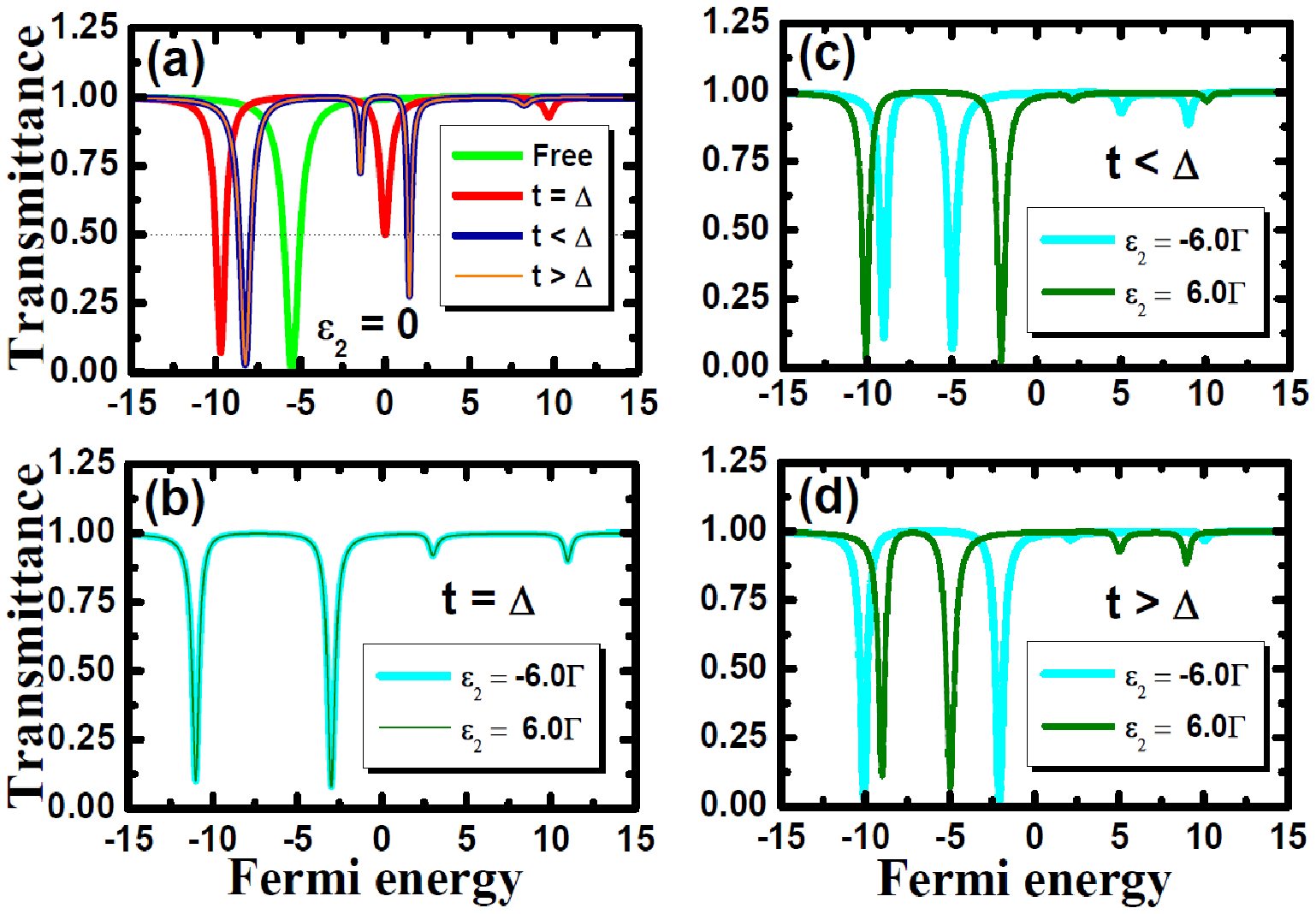}
\protect\protect\protect\protect\protect\protect\protect\protect\protect\protect\protect\protect\protect\protect\protect\protect\protect\protect\protect\protect\protect\protect\protect\protect\protect\protect\protect\protect\protect\protect\protect\caption{\label{Fig3}(Color online) Transmittance as a function of the Fermi
energy of the STM tips within the Fano regime $x=1$ $(q_{b}=0)$:
panels (a)-(d) display the same features of Fig. \ref{Fig2} in the
opposite regime of interference.}
\end{figure}

Fig.\,\ref{Fig3} holds within the Fano limit $x=1$ $(q_{b}=0)$
where the electron tunneling between the STM tips is the dominant
process in the system, thus resulting in Fano antiresonances instead
of peaks in the transmittance profiles as a function of the Fermi
energy. Panel (a) first displays the case in which the adatom 1 is
decoupled from the adatom 2 represented by the green curve characterized
by a dip at $\varepsilon=\varepsilon_{1}=-5$. By using finite values
for $t$ and $\Delta$ combined with $\varepsilon_{2}=0,$ we can
observe the crossover from the regime $t\neq\Delta$ ($t=2$ with
$\Delta=4$ for the blue lineshape and $t=4$ with $\Delta=2$ in
the case of the curve for the orange color) towards the point $t=\Delta=4,$
where we can clearly realize in the red curve the emergence of a dip with amplitude $\mathcal{T}=1/2,$ analogously to the opposite Fano regime of interference ($x=0$ and $q_{b}\rightarrow\infty$) found in Fig.\,\ref{Fig2}(a). In presence of the potential $\varepsilon_{2}\neq0,$ the zero-bias
dip disappears according to the curves with $\varepsilon_{2}=6\Gamma$
(positive potential) and $\varepsilon_{2}=-6\Gamma$ (negative potential)
as found in panel (b) of the same figure. As in Fig.\,\ref{Fig2}(b)
we also report a universality feature in the transmittance profile,
which still arises from the condition $\Sigma_{\text{{MFs}}}(\varepsilon_{2})=\Sigma_{\text{{MFs}}}(-\varepsilon_{2})$
apart from the Fano parameter as we can notice in Eq.\,(\ref{eq:MQPself}).
For $t\neq\Delta$ coincident curves no longer exist and the universal
behavior is not verified as pointed out by panels (c) and (d), which
have the same set of parameters as in Fig.\,\ref{Fig2}.

\begin{figure}[!htb]
\includegraphics[width=0.5\textwidth,height=0.3\textheight]{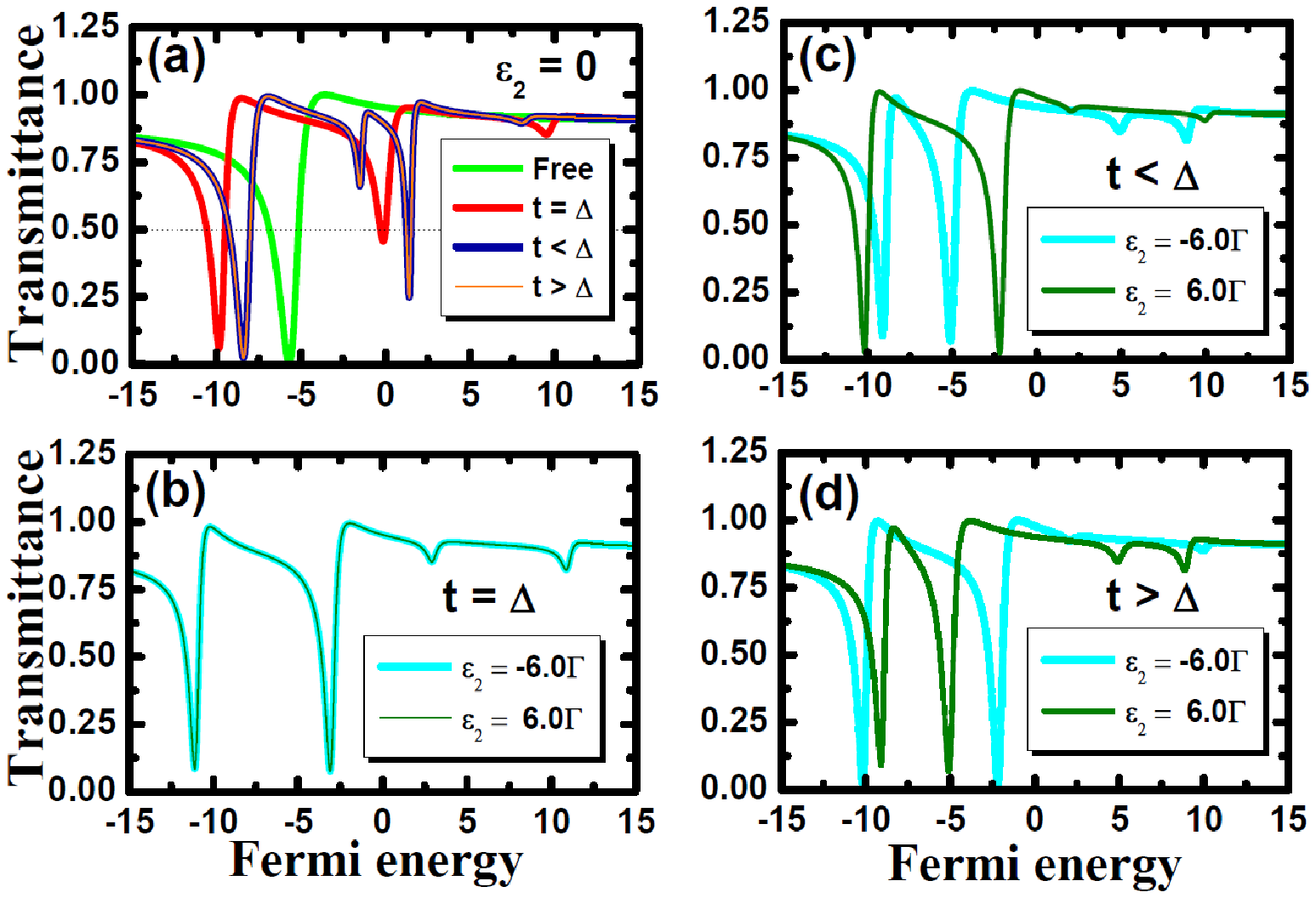}
\protect\protect\protect\protect\protect\protect\protect\protect\protect\protect\protect\protect\protect\protect\protect\protect\protect\protect\protect\protect\protect\protect\protect\protect\protect\protect\protect\protect\protect\protect\protect\caption{\label{Fig4}(Color online) Transmittance as a function of the Fermi
energy of the STM tips within the Fano regime $x=0.5$ $(q_{b}\approx0.35)$:
panels (a)-(d) display the same features of Figs.\ref{Fig2} and \ref{Fig3},
thus attesting that the point $t=\Delta$ is protected against the
Fano effect as well as the symmetric change of the level in the adatom
2.}
\end{figure}

\begin{figure}[!htb]
\includegraphics[width=0.5\textwidth,height=0.3\textheight]{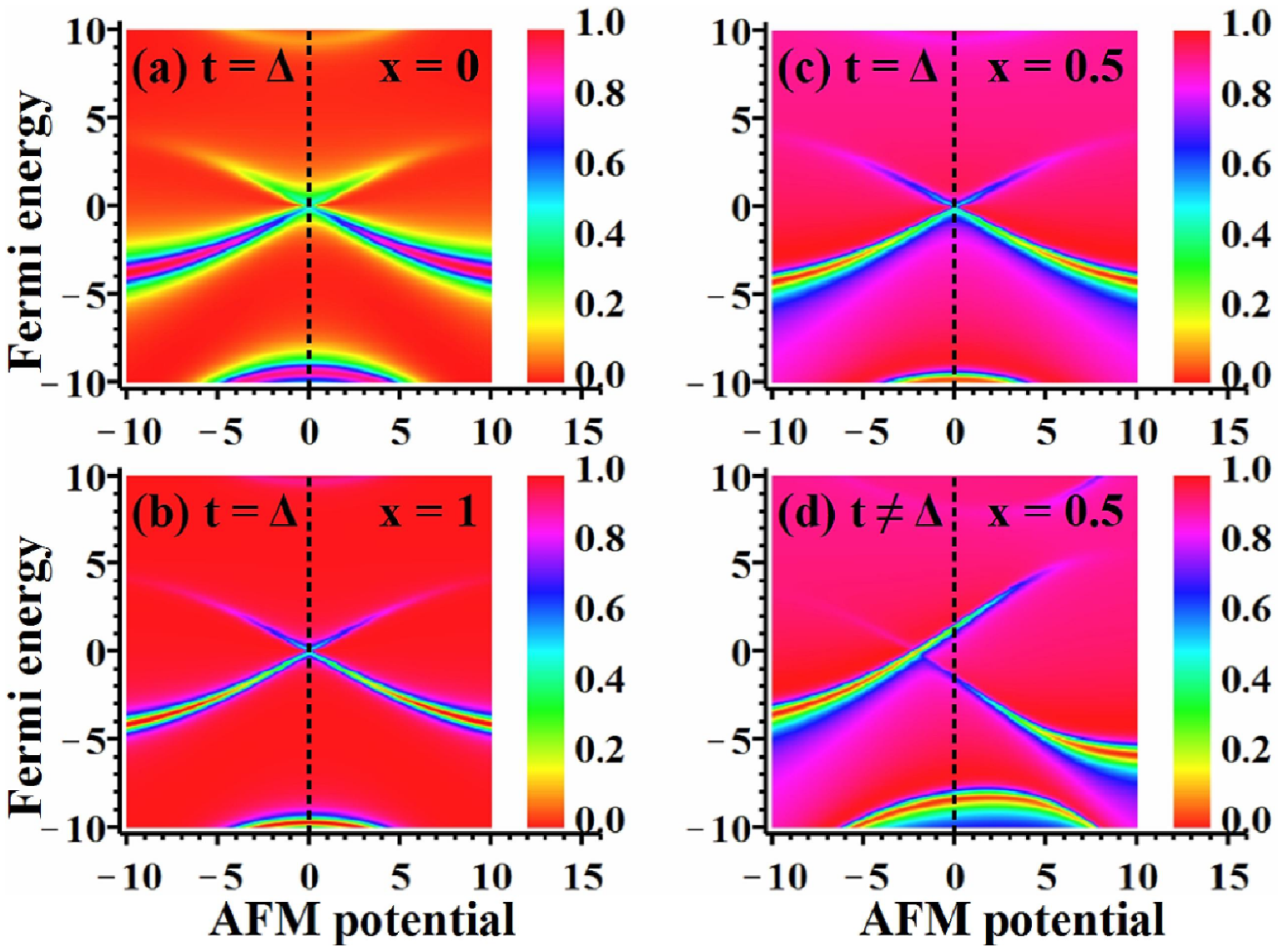}
\protect\protect\protect\protect\protect\protect\protect\protect\protect\protect\protect\protect\protect\protect\protect\protect\protect\protect\protect\protect\protect\protect\protect\protect\protect\caption{\label{Fig5}(Color online) Density plots for the transmittance as
a function of the Fermi energy of the STM tips and the potential $\varepsilon_{2}$
tuned by the AFM tip in distinct Fano limits: (a) $x=0$ $(q_{b}\rightarrow\infty)$,
(b) $x=1$ $(q_{b}=0)$, (c) $x=0.5$ $(q_{b}\approx0.35)$ for $t=\Delta$
and (d) $x=0.5$ $(q_{b}\approx0.35)$ in the situation $t\protect\neq\Delta.$
The symmetrical panels (a), (b) and (c) suggest that a pair of MFs
is formed in which only the MF $A$ couples to the adatom 1 (see Fig.\ref{Fig1}(d)).
In panel (d), the absence of the mirror symmetry under analysis arises
from the simultaneous coupling of MFs $A$ and $B$ with the adatom
1. It occurs via the distinct amplitudes $(t+\Delta)$ and $(\Delta-t)$
as Eq. (\ref{eq:MQPs}) ensures for $t\protect\neq\Delta.$}
\end{figure}

To make explicit that the invariance feature of the gate potential
for the point $t=\Delta=4$ is achievable for any Fano ratio $q_{b},$
we present in Fig.\,\ref{Fig4} the case $x=0.5$ $(q_{b}\approx0.35)$
in which both paths $V$ and $V_{12}$ of Eq.\,(\ref{eq:TIAM}) compete
on an equal footing. For this situation, we find intermediate Fano
profiles where the underlying physics of Figs.\,\ref{Fig2} and \ref{Fig3} is still the same. Therefore based on the results of Figs.$\ref{Fig2},$
\ref{Fig3} and \ref{Fig4}, we demonstrate that the invariance of
the Fano profile is independent on $q_{b}$ arising solely from the
feature $\Sigma_{\text{{MFs}}}(\varepsilon_{2})=\Sigma_{\text{{MFs}}}(-\varepsilon_{2}),$
due to the connected MFs appearing in the term $i\varepsilon_{2}\Psi_{A}\Psi_{B}$
within Eq. (\ref{eq:MQPs}), in particular for $t=\Delta.$ Moreover,
the invariance with the potential $\varepsilon_{2}$ in the transmittance
becomes clearer if we look to its density plot spanned by the axes
$\varepsilon$ (Fermi level) and $\varepsilon_{2}$. Fig.\,\ref{Fig5}(a)
is for $\varepsilon_{1}=-5$ and $t=\Delta=4$: it exhibits the case
$x=0$ $(q_{b}\rightarrow\infty)$ for the regime of Fano interference,
which shows the mirror symmetry under consideration with respect to
the vertical axis placed at $\varepsilon_{2}=0$ (see the vertical
dashed lines in the same figure). Notice that such a feature also
manifests itself in panels (b) and (c), respectively in the limits
$x=1$ and $x=0.5.$ Here the orange color designates perfect insulating
regions and those conducting are represented by red color. In panel
(d) of the current figure, this mirror symmetry signature is broken
just by using $t\neq\Delta$ as expected $(\Delta=2$ and $t=4$).

The invariant Fano profiles for the transmittance found in panels
(b) of the Figs.\,\ref{Fig2}, \ref{Fig3} and \ref{Fig4} as well
as those (a), (b) and (c) for Fig.\,\ref{Fig5} are due to the symmetric
swap of the adatom 2 level around the MF zero mode which reveals the
formation of a pair of MFs within such an adatom when $t=\Delta.$
It couples the MF $A$ to the adatom 1 with amplitude $\sqrt{2}\Delta$
as Eq.(\ref{eq:MQPs}) ensures. However in the regime $t\neq\Delta,$
the pair of MFs still exists within the adatom 2, but with the MFs
$A$ and $B$ hybridized distinctly with the adatom 1 via the strengths
$(t+\Delta)$ and $(\Delta-t),$ respectively. This feature is then
probed by the symmetric tuning of the potential $\varepsilon_{2}$
around the MF zero mode, which yields the panels (c)-(d) of the Figs.\,\ref{Fig2},
\ref{Fig3} and \ref{Fig4}, in addition to the density plot in panel
(d) of Fig.\,\ref{Fig5}. Therefore, the aforementioned mechanism
ruling the transmittance profiles via the couplings of the MFs with
the adatom 1 is encoded by the self-energy $\Sigma_{\text{{MFs}}}$
of Eq.\,(\ref{eq:MQPself}).

\section{Conclusions}

\label{sec4}

In summary, we have explored theoretically a Fano interferometer composed
by STM and AFM tips over superconducting adatoms, in which the pair
of MFs under the latter, elucidates the gate invariance feature of
the interferometer, due to the Majorana nature arising from the adatom
2. Particularly for the situation where only one MF hybridizes with
the adatom coupled to STM tips, the aforementioned invariance consists
of a universal behavior within the transmittance when the AFM tip
tunes symmetrically the energy level of its adatom around the MF zero
mode. Such a universality is constituted by a common Fano profile
in transmittance as a function of the Fermi level of the STM tips
for two symmetric values for the AFM tip potential. In the case of
two MFs connected to the adatom beneath the STM tips, we verify that
such a universality is broken. Hence, despite the experimental challenging
of the proposal, we expect that in the near future such an interferometer
can be developed.

\section*{Acknowledgments}

This work was supported by the Brazilian agencies CNPq, CAPES,
$2014/14143-0$ and $2015/23539-8$ São Paulo Research Foundation (FAPESP).

\appendix

\section{Transmittance derivation}

Here we use the interaction picture to calculate $\mathcal{T}(\varepsilon)$
from Eq.(\ref{eq:transmit-1}). It ensures that a state $\left|\Phi_{n}\right\rangle $
from the spectrum of the Hamiltonian given by $\mathcal{H}_{e}+\mathcal{H}_{o}=\mathcal{H}_{\varphi=0}$
admits the following time-dependency

\begin{align}
\left|\Phi_{n}\right\rangle  & =e^{-\frac{i}{\hbar}\int_{-\infty}^{0}\mathcal{\tilde{H}}_{\text{{tun}}}(\tau)d\tau}\left|\Psi_{n}\right\rangle \nonumber \\
 & \simeq(1-\frac{i}{\hbar}\int_{-\infty}^{0}\mathcal{\tilde{H}}_{\text{{tun}}}(\tau)d\tau)\left|\Psi_{n}\right\rangle ,\label{eq:phi}
\end{align}
where $\hbar=\frac{h}{2\pi}$ and $\left|\Psi_{n}\right\rangle $
is an eigenstate of $\mathcal{H}_{\varphi=0}.$ Thus the current $\mathcal{J}_{\text{{tip-1}}}$
for the STM tip 1 can be obtained by performing the expected mean
value of the current operator $\mathcal{I}_{\text{{tip-1}}}\equiv\mathcal{I}_{\text{{tip-1}}}\left(t=0\right)$,
which reads

\begin{align}
\mathcal{J}_{\text{{tip-1}}} & =\left\langle \Phi_{n}\right|\mathcal{I}_{\text{{tip-1}}}\left|\Phi_{n}\right\rangle \nonumber \\
 & =-\frac{i}{\hbar}\left\langle \Psi_{n}\right|\int_{-\infty}^{0}[\mathcal{I}_{\text{{tip-1}}},\mathcal{\tilde{H}}_{\text{{tun}}}(\tau)]d\tau\left|\Psi_{n}\right\rangle\nonumber\\
 & +\mathcal{O}(\mathcal{\tilde{H}}_{\text{{tun}}}^{2}),\label{eq:current}
\end{align}
where we have regarded $\left\langle \Psi_{n}\right|\mathcal{I}_{\text{{tip-1}}}\left|\Psi_{n}\right\rangle =0$
and by considering the thermal average on the latter equation, which gives

\begin{align}
\mathcal{J}_{\text{{tip-1}}}=-\frac{i}{\hbar}\int_{-\infty}^{0}{\tt Tr}\{\varrho_{\varphi=0}[\mathcal{I}_{\text{{tip-1}}},\mathcal{\tilde{H}}_{\text{{tun}}}(\tau)]\}d\tau,\nonumber \\
\label{eq:_07c}
\end{align}
where $\varrho_{\varphi=0}$ is the density matrix of the system described
by the Hamiltonian $\mathcal{H}_{\varphi=0}.$ By applying the equation-of-motion
on $\mathcal{I}_{\text{{tip-1}}}$, we show that

\begin{eqnarray}
\mathcal{I}_{\text{{tip-1}}} & = & -\frac{i}{\hbar}[e\sum_{k}c_{1k}^{\dagger}c_{1k},\mathcal{H}_{\varphi=0}]\nonumber \\
 & = & \left(-\frac{ie}{\sqrt{2}\hbar}\right)V\sum_{k}\left\{ (c_{ek}^{\dagger}d_{1}-d_{1}^{\dagger}c_{ek})\right.\nonumber \\
 & + & \left.(c_{ok}^{\dagger}d_{1}-d_{1}^{\dagger}c_{ok})\right\} \nonumber \\
 & + & \left(-\frac{ie}{\hbar}\right)V_{12}\sum_{q\tilde{q}}(c_{oq}^{\dagger}c_{e\tilde{q}}-c_{e\tilde{q}}^{\dagger}c_{oq}),\nonumber \\
 & &\label{eq:11b}
\end{eqnarray}
which, in combination with Eq. (\ref{eq:_07c}), leads to
\begin{align}
\mathcal{J}_{\text{{tip-1}}} & =-\frac{e}{\hbar}\Delta\mu{\tt Im}\int_{-\infty}^{+\infty}d\tau\{\sqrt{2}V\mathcal{F}(-\tau)\nonumber \\
 & +2V_{12}\mathcal{M}(-\tau)\},\label{eq:J_B}
\end{align}
where

\begin{align}
\mathcal{F}(-\tau) & =-\frac{i}{\hbar}\theta(-\tau){\tt Tr}\{\varrho_{\varphi=0}[f_{o}^{\dagger}d_{1},\sum_{k}c_{ek}^{\dagger}(\tau)c_{ok}(\tau)]\}\label{eq:F}
\end{align}
and

\begin{align}
\mathcal{M}(-\tau) & =-\frac{i}{\hbar}\theta(-\tau){\tt Tr}\{\varrho_{\varphi=0}[f_{o}^{\dagger}f_{e},\sum_{k}c_{ek}^{\dagger}(\tau)c_{ok}(\tau)]\}\label{eq:M}
\end{align}
are retarded Green's functions.

In order to find a closed expression for the current $\mathcal{J}_{\text{{tip-1}}}$,
we should evaluate the integrals in the time coordinate $\tau$ of
Eq. (\ref{eq:J_B}), which result in

\begin{align}
\int_{-\infty}^{+\infty}d\tau\mathcal{F}(-\tau)& =  \mathbb{\mathcal{Z}}^{-1}\sum_{mn}\frac{(e^{-\beta E_{n}}-e^{-\beta E_{m}})}{E_{n}-E_{m}+i0^{+}}\nonumber \\
 & \times  \left\langle \Psi_{n}\right|f_{o}^{\dagger}d_{1}\left|\Psi_{m}\right\rangle\nonumber \\
 & \times  \left\langle \Psi_{m}\right|\sum_{k}c_{ek}^{\dagger}c_{ok}\left|\Psi_{n}\right\rangle \label{eq:int_1}
\end{align}
and

\begin{align}
\int_{-\infty}^{+\infty}d\tau\mathcal{\mathcal{M}}(-\tau) & =  \mathbb{\mathcal{Z}}^{-1}\sum_{mn}\frac{(e^{-\beta E_{n}}-e^{-\beta E_{m}})}{E_{n}-E_{m}+i0^{+}}\nonumber \\
 & \times  \left\langle \Psi_{n}\right|f_{o}^{\dagger}f_{e}\left|\Psi_{m}\right\rangle \nonumber \\
 & \times  \left\langle \Psi_{m}\right|\sum_{k}c_{ek}^{\dagger}c_{ok}\left|\Psi_{n}\right\rangle ,\label{eq:int_2}
\end{align}
where we have used $\mathbb{\mathcal{Z}}$ as the partition function
of $\mathcal{H}_{\varphi=0}\left|\Psi_{m}\right\rangle =E_{m}\left|\Psi_{m}\right\rangle $,
$\mathcal{A}\left(\tau\right)=e^{\frac{i}{\hbar}\mathcal{H}_{\varphi=0}\tau}\mathcal{A}e^{-\frac{i}{\hbar}\mathcal{H}_{\varphi=0}\tau}$
for an arbitrary time-dependent operator $\mathcal{A}\left(\tau\right).$
To eliminate the matrix element $\left\langle \Psi_{m}\right|c_{ek}^{\dagger}c_{ok}\left|\Psi_{n}\right\rangle $
in Eqs. (\ref{eq:int_1}) and (\ref{eq:int_2}), we calculate $\left\langle \Psi_{m}\right|[\sum_{k}c_{ek}^{\dagger}c_{ok},\mathcal{H}_{\varphi=0}]\left|\Psi_{n}\right\rangle $,
which gives

\begin{align}
\left\langle \Psi_{m}\right|\sum_{k}c_{ek}^{\dagger}c_{ok}\left|\Psi_{n}\right\rangle  & =  \frac{-\sqrt{2}V\left\langle \Psi_{m}\right|d_{1}^{\dagger}f_{o}\left|\Psi_{n}\right\rangle }{(E_{n}-E_{m})}\nonumber \\
 & -  \frac{2V_{12}\left\langle \Psi_{m}\right|f_{e}^{\dagger}f_{o}\left|\Psi_{n}\right\rangle }{(E_{n}-E_{m})}.\nonumber \\
\label{eq:mix}
\end{align}

By performing the substitutions of Eqs. (\ref{eq:int_1}), (\ref{eq:int_2})
with (\ref{eq:mix}) in Eq. (\ref{eq:J_B}), we enclose the result
into the function labeled by $\chi_{mn}$ to show that
\begin{align}
\mathcal{J}_{\text{{tip-1}}} & =\frac{e}{\hbar}\pi\Delta\mu\mathbb{\mathcal{Z}}^{-1}\sum_{mn}\chi_{mn}\frac{(e^{-\beta E_{n}}-e^{-\beta E_{m}})}{E_{n}-E_{m}}\nonumber \\
 & \times\delta(E_{n}-E_{m})\nonumber \\
 & =-\frac{e}{\hbar}\pi\Delta\mu\beta\sum_{mn}[\mathbb{\mathcal{Z}}^{-1}e^{-\beta E_{n}}\delta(E_{n}-E_{m})]\nonumber \\
 & \times\chi_{nm},\label{eq:J_B_2}
\end{align}
where we have defined
\begin{multline}
\chi_{nm}=(\sqrt{2}V)^{2}\left\langle \Psi_{n}\right|f_{o}^{\dagger}d_{1}\left|\Psi_{m}\right\rangle \left\langle \Psi_{m}\right|d_{1}^{\dagger}f_{o}\left|\Psi_{n}\right\rangle \\
+2\sqrt{2}V(2V_{12})\left\langle \Psi_{n}\right|f_{o}^{\dagger}d_{1}\left|\Psi_{m}\right\rangle \left\langle \Psi_{m}\right|f_{e}^{\dagger}f_{o}\left|\Psi_{n}\right\rangle \\
+(2V_{12})^{2}\left\langle \Psi_{n}\right|f_{o}^{\dagger}f_{e}\left|\Psi_{m}\right\rangle \left\langle \Psi_{m}\right|f_{e}^{\dagger}f_{o}\left|\Psi_{n}\right\rangle .\label{eq:qui_mn}
\end{multline}

In this calculation we have used
\begin{multline*}
\left\langle \Psi_{n}\right|f_{o}^{\dagger}d_{1}\left|\Psi_{m}\right\rangle \left\langle \Psi_{m}\right|f_{e}^{\dagger}f_{o}\left|\Psi_{n}\right\rangle \\
=\left\langle \Psi_{n}\right|f_{o}^{\dagger}f_{e}\left|\Psi_{m}\right\rangle \left\langle \Psi_{m}\right|d_{1}^{\dagger}f_{o}\left|\Psi_{n}\right\rangle ,
\end{multline*}
with
\begin{equation}
\frac{(e^{-\beta E_{n}}-e^{-\beta E_{m}})}{E_{n}-E_{m}}=-\beta e^{-\beta E_{n}}\label{eq:_limit}
\end{equation}
in the limit $E_{n}\rightarrow E_{m}$. The property $\left[\mathcal{H}_{e},\mathcal{H}_{o}\right]=0$
ensures the partitions $E_{n}=E_{n}^{e}+E_{n}^{o}$ and $\mathbb{\mathcal{Z}}=\mathbb{\mathcal{Z}}_{e}\mathbb{\mathcal{Z}}_{o}$
for the Hamiltonians $\mathcal{H}_{e}$ and $\mathcal{H}_{o}$, respectively
in the brackets of Eq. (\ref{eq:J_B_2}), thus leading to

\begin{align}
 & \mathbb{\mathcal{Z}}^{-1}e^{-\beta E_{n}}\delta(E_{n}-E_{m})=\frac{1}{\beta}\mathbb{\mathcal{Z}}_{e}^{-1}\mathbb{\mathcal{Z}}_{o}^{-1}\int d\varepsilon\left(-\frac{\partial f_{F}}{\partial\varepsilon}\right)\nonumber \\
 & \times(e^{-\beta E_{n}^{e}}+e^{-\beta E_{m}^{e}})(e^{-\beta E_{n}^{o}}+e^{-\beta E_{m}^{o}})\nonumber \\
 & \times\delta(\varepsilon+E_{n}^{e}-E_{m}^{e})\delta(\varepsilon+E_{n}^{o}-E_{m}^{o}).\label{eq:prop}
\end{align}

Therefore, we substitute Eqs. (\ref{eq:qui_mn}) and (\ref{eq:prop})
in Eq. (\ref{eq:J_B_2}) to calculate $G=\frac{\partial}{\partial\varphi}\mathcal{J}_{\text{{tip-1}}}(\varphi=0)$.
The comparison of such a result with Eq. (\ref{eq:_10b}) allows us
to find Eq.(\ref{eq:transmit-1}). We also verify that $G=\frac{\partial}{\partial\varphi}\mathcal{J}_{\text{{tip-2}}}(\varphi=0)$
for the STM tip 2, which is agreement with Refs.\citep{key-371,Zheng}
that show for the case with symmetric couplings $V$ absence of Andreev
currents, i.e., $\frac{\partial}{\partial\varphi}\mathcal{J}_{\text{{tip-1}}}(\varphi=0)\neq\frac{\partial}{\partial\varphi}\mathcal{J}_{\text{{tip-2}}}(\varphi=0).$

\end{document}